\documentclass[manuscript]{aastex}

\slugcomment{submitted to Astrophysical Journal Letters}

\shorttitle{SN 2014J Light Echoes}
\shortauthors{Crotts}

\begin{document}

\title{Light Echoes From Supernova 2014J in M82}

\author{Arlin P.S. Crotts}
\affil{Department of Astronomy, Columbia University, 550 West 120th St.,
New York, NY~ 10027}
\email{arlin@astro.columbia.edu}

\begin{abstract}
Type Ia SN 2014J exploded in the nearby starburst galaxy M82 = NGC 3032, and
was discovered at Earth about seven days later on 2014 January 21, reaching V
maximum light around 2014 February 5.
SN 2014J is the closest SN Ia in at least four decades and probably many more.
Recent $HST$/WFC3 imaging (2014 September 5 and 2015 February 2) of M82 around
SN 2014J reveals a light echo at radii of about 0.6 arcsec from the SN
(corresponding to about 12 pc at the distance of M82).
Likely additional light echoes reside at a smaller radii of about 0.4 arcsec.
The major echo signal corresponds to echoing material about 330 pc in the
foreground of SN 2014J, and tends to be bright where pre-existing nebular
structure in M82 is also bright.
The second, likely echo corresponds to foreground distances of 80 pc in front
of the SN.
Even one year after maximum light, there are indications of further echo
structures appearing at smalle radii, and future observations may show how 
extinction in these affect detected echo farther from the SN, which will
affect interpretion of details of the three-dimensional structure of this gas
and dust.
Given enough data we might even use these considerations to constrain the
near-SN material's shadowing on distant echoing clouds, even without directly
observing the foreground structure.
This is in addition to echoes in the near future might also reveal
circumstellar structure around SN 2014J's progenitor star from direct imaging
observations and other techniques.
\end{abstract}


\section{Introduction}

On January 14, 2014, SN 2014J flared into view in M82 \citep{zhe14},
to be discovered on January 21/22 \citep{fos14},
perhaps the closest SN Ia since SN 1885 in M31,\footnote
{SN 2014J in M82 = NGC 3034 is at a distance of $3.5\pm 0.3$ Mpc, while SN 1885
in M31 was 0.8 Mpc away.
SN 1937C in IC 4182 was $4.0\pm 0.5$ Mpc away, SN 1986G in Cen A was
$3.9\pm 1.0$ Mpc distant, while SNe 1895B and 1972E in NGC 5253 were at
$3.9 \pm 0.7$ Mpc.}
and the closest SN of any type observed since SN 1987A.
SN 2014J is also special for its appearance in the highly active starburst
galaxy M82 (0.9 kpc from its center), but being a SN Ia this supernova will
sample a region of space in M82 that is not pre-determind to include a
star-formation region, as in the case of a core-collapse SN.

Several studies list the extinction to SN 2014J variously as
$A_V=1.85 \pm 0.11$ \citep{ama14},
$\sim$2$-3$ \citep{bro14},
and 3.14$\pm$0.11 \citep{fol14},
yielding $A_V = 2.6$ mag when averaged linearly in flux, translating to
$A_B \approx 4$ in the other band we consider primarily here (see \citet{goo14},
\citet{fol14}).

\section{Observations}

These observations result from a $HST$/WFC3 program (\#13626: Crotts, PI) to
observe properties of the light echoes and progenitor environment around
SN 2014J.
They consisted of five series of short exposures, primarily in single-orbit
visits, with the idea of finding increasing deeper imaging structure as the SN
fades.
In the latest two of these five epochs the SN was sufficiently faint to reveal
the echo signals discussed here without being swamped by the SN source itself.
All five visits will be useful for further investigations to be discussed in
later work.
The primary observations used in this paper are a total of 576s exposure on 
2014 September 5 in the F438W filter, 560s exposure in the F555W filter, and
512s in F814W, and on 2015 February 2 with 1536s in F438W and 384s in F555W.
More accurately, these were taken on UT 2014 September 5.9 (= JD 2456906.4 =
MJD 56905.9 = 234.2 days after the estimated appearance of SN 2014J on 2014
Jan 14.75 = MJD 56671.75 and 213 days after maximum in V), and with the later
epoch on UT 2015
February 2.6, or 149.7 days later.
The point-spread function for each of these two bands is derived from an 8s
exposure in F814W and 40s in F438W on 56727.8 (day 56.0 after outburst), and an
128s exposure in F555W on 56781.1 (day 109.6).
As a point of reference, our photometry of SN 2014J on day 234.2 in F438W,
F555W and F814W (STMAG = 16.70, 16.62 and 16.83, respectively) transforms
roughly to $(B,V,I)$ values of 16.9, 16.7, and 16.8, with $B$ more uncertain.

\section{Analysis}

A light echo at uniform distance well in the SN foreground will resemble a ring
or arc of light of a constant radius of curvature.
That ring or arc will appear centered on the SN, unless the sheet of reflecting
material is tilted versus the sightline from the observer to the SN, in which
case it will appear as a ring/arc off-center from the SN.
Any echo, therefore, is usually composed of a composite of rings or arcs, even
in the case of the SN imbedded in reflecting nebulosity, in which case these
rings/arcs can extend to zero angular radius, in an extended fuzz of
illumination.
Because of these characteristics, echoes have a strong tendency to appear as
arcs/rings centered near the SN, unless they are at small angular radii.
``Small'' angular radii in this case are on the scale of $ct$ at the distance
of M82, where $c$ is the speed of light and $t$ is the time since the light
pulse maximum (213d for these observations).
At the distance of M82, this corresponds to 0.023 arcsec diameter, 58\% the
width of a WFC3/UVIS pixel, hence unresolved, and inaccessible for faint
surface brightnesses, due to the bright point source of the SN.

The foreground distance $z$ of echoing material is approximated by the
expression $z = r^2/2ct - ct/2$, where $r$ is the physical distance transverse
to the Earth-SN sightline, to the echo's position.
This equation for a paraboloid is an accurate approximation for the ellipsoid
with one focus at Earth and one focus at the SN, with a major axis longer than
the Earth-SN distance by $ct$.
One notable characteristic of echoes is that for foreground distance $z >> ct$,
and for a sheet of material even roughly perpendicular to the Earth-SN
sightline, the apparent transverse motion of the echo is almost always faster
than lightspeed, hence a reliable signature for the presence of an echo.

Fig.\ 1 shows several aspects of the field around the SN 2014J seen in the
band F438W (for WFC3/UVIS images) and F435W for ACS/WFC and all for the same
field of view, 8.4 arcsec = 143 pc across.
Fig.\ 1a shows the field taken by ACS/WFC for a Hubble Heritage image on 2006
March 29 (program \#10776: Mountain, PI), 2848 d before SN 2014J's first light.
This view of M82, 0.9 kpc west of the galaxy's nucleus, is centered on SN 2014J
indicated by the circular mark 10 pc diameter, centered on the SN.
Note that the SN is centered at one end of a dark lane, and sits just west of a
particularly bright patch of nebulosity.

Fig.\ 1b shows the same 143 pc field in M82 as in Fig.~1a, taken by
$HST$/WFC3(UVIS) on 2014 September 5 (213 d after maximum) in F438W as part of
program \#13626 (Crotts, PI).
SN 2014J is seen as the bright point source in the center.
Note in addition the apparent nebulosity out to radii of about 13 pc,
especially just to the east of the SN.
(The nebulosity of M82 is less apparent than in Fig.~1a due to the reduced
contrast in displaying these data.)
Fig.\ 2a shows the image in Fig.\ 1a subtracted from that in Fig.\ 1b, after
being scaled in flux and registered in spatial coordinates.
The background does not subtract perfectly, leaving a mottled appearance due to
color terms left from the F435W band in ACS(WFC) versus the F438W band in
WFC3(UVIS); 88\% of transmissivity in F435W and 98\% in F438W are contained in
the overlap between the two bands.
Note that even more apparent residual nebulosity has arisen with the presence
of SN 2014J, at radii of about 0.6 arcsec, faint but evident.

Fig.\ 2b corresponds to Fig.\ 2a with the SN point source subtracted (as
derived from a 8s F438W WFC3/UVIS SN image from 35 d after maximum).
The echo ring at 11 - 13 pc is even more apparent, mainly south of the SN (at
PAs 85 to 230), and to a lesser extent due North (PA -40 to 30), but not
significantly to the East and West (PA 50 to 85, and 230-250 and 275-315,
respectively).
The echoes are absent from the area noted as the dark lane in Fig.\ 1a, while
they correlate well with bright nebulosity at other PAs.
The position of the echo was measured by centroiding the signal in cross-cuts
across the rings every 5 degrees in PA (1 pixel width at the echo radius), with
errors estimated from the dispersion of surrounding pixel values, accounting
for the number of pixels accross the echo peak.
In addition to the prominent echo ellipse at about 0.6 arcsec radius, we search
for an other significant, transient bright spots and find one plausible echo
candidate at smaller radius $r \approx 0.4$ arcsec, PA$\approx 215^\circ$.
This is only a 4-sigma deviation given the range of PSF-subtraction residuals
at this small radius, but we detail this feature shortly.

We have the opportunity to view this same field observed five months later in
the same bands, and look for arc-like structures moving at apparent
superluminal speeds.
These are shown most readily by subtracting the 2014 September epoch from the
2015 February epoch, which will readily reveal brighter echoes as a
positive signal in advance of a negative one.
New echoes will appear as positive-only signals.
This can be seen in Fig.\ 3 for F438W.
Where the 11 - 13 pc ring was strongest in 2014 September e.g., PA 85 to 170,
it is still strong.
At other PAs in this representation it is marginally detected.
Significantly, the hint of an inner echo at $r \approx 0.4$ arcsec, PA$\approx
215^\circ$ is confirmed; in fact this signal has spread to PA 205 to 225.
Additionally, strong positive-only signals have appeared at similar radii
$0.3$ arcsec $\la r \la 0.5$ arcsec, beyond the range of PSF
subtraction systematic errors, suggesting a complex of structures ranging over
about a factor of $\pm 40$\% in foreground distance $z$.

Fig.\ 4 shows the derived three-dimensional geometry of the echoing material
seen in the first epoch and confirmed in the second, showing an extensive
structure about 330 pc in the SN foreground, with the additional, possible
structure at about 80 pc.
The outer arcs have notable slopes in foreground distance across the field (the
southern structure appearing farther from the SN in its southern extremes, and
the northern structure more perpendicular to the sightline).
At the distance of SN 2014J from the galaxy's center 0.9 kpc out on the major
axis, most of the gas and dust there appears within 1 kpc of the axis
(ignoring that M82 might be edge-on).
The observation that SN 2014J is at least 330 pc behind major structure
(probably coincident with prominent luminous nebulosity seen in Fig.\ 1a)
indicates it is deep within M82 as seen from Earth.
These loci are derived as shown in Fig.\ 5a with the centroids and errorbars of
the radius from the SN of the primary echo around SN 2014J as a function of
position angle, as of 213 d after maximum light, plus a best-fit ellipse
centered on the SN point source.
Centroids were made by taking radial (or roughly radial) cross-cuts of pixels,
where the echo width along each cross-cut was estimated separately, but
corresponding to 2 to 4 pixels (hence about 3 to 6 months of echo motion).
This is fit with a SN-centered ellipse of nearly north-south orientation and an
ellipticity of 0.20.

We used the echo locus in Fig.\ 4 to measure both the surface brightness of the
echo on day 213 after maximum in F438W and the corresponding surface brightness
in F438W (in ACS) from 2006, before the SN.
These are shown in Fig.\ 5b: the surface brightness at the echo locus of the
underlying M82 nebulosity (small crosses) in the ACS/F435W band and light echo
(large dots) in the WFC3/F438W band.
The echo surface brightness is not determined (but consistent with zero) at
certain position angles where the width of the echo cannot be measured, as
indicated by bars at zero surface brightness near PA = 50, 240 and 300$^\circ$.
Surface brightnesses are in units of approximately $10^{-18}$ erg s$^{-1}$
cm$^{-2}$ \AA$^{-1}$ arcsec$^{-2}$.
We note that the echo surface brightness is well-correlated (if not perfectly
so) with the underlying brightness of nebulosity, and discuss the possible
significance of this below.

SN2014J's lightcurve was observed at and near maximum light (\citet{goo14},
\citet{kaw14}, \citet{mar14}, \citet{tsv14}, \citet{zhe14}), including studies
using the same WFC3 bands as in this study (\citet{ama14}, \citet{fol14}).
Integrating the fluence over the maximum light peak (within 2 mag of the peak
in B), produces m(F438W)$ - $m(F555W) $=$ 1.15, whereas the F438W/F555W color
of the echo itself is 0.75, about 0.4 mag bluer, hence with a wavelength
dependence in scattering efficiency of $Q_{scat} \approx \lambda^{-1.5}$.
This is similar to the echo photometry from SN 1987A, in which the echoes show
$B-V = 1.1$ to 1.2 e.g., \citep{sun88}, whereas the maximum light colors of
SN 1987A were 1.6 (\citet{ham88}, \citet{men87}, \citet{cat88}), also 0.4 mag
redder than its echoes.

\section{Discussion}

Beyond its geometry, interpretation of the echoes is in large part dependent on
the optical depth ($\tau_{sc}$ of the scattering dust, with multiple
scattering beginning to dominate for $\tau_{sc} > 1$.
As cited above, estimates toward SN 2014J itself vary from $\tau_{sc} = 1.7$ to
2.9 in V (or F555W), hence up to $\tau_{sc} \approx 3.7$ in B (or F438W).
Are large optical depths borne out by the behavior of the echoes themselves?

The echo-derived estimates of $Q_{scat}$, $\tau_{sc}$ and other properties will
depend on the assumption that the extinction and scattering along the direct
sightline from SN to Earth is identical to the reflected path from SN to
echoing material to Earth, a deviation of only $2^\circ$.
However, the environment around the SN is complex, and the SN resides in a dark
lane (in projection) while the echoing material does not.
Furthermore, while the echo brightness correlates with the brightness of
nebular emission, this not a perfect indicator (Fig.\ 5b).

Nonetheless, Patat (2005) argues that the average properties of an echo complex
can be correlated with the brightness of the SN itself and its extinction
environment.
Patat calculates (his Figure 6) the ratio of echo brightness to peak SN
brightness, for a foreground echo sheet, which is strongly a function of
$\tau_{sc}$ but very weakly a function of time since explosion for the first
decades of echoes.
Is SN 2014J's echo similar in behavior for such large optical depths to the
(small) sample of other observed SN Ia echoes, which tend to be dominated by
foreground material (versus sometimes material in the SN vicinity for core
collapse SN)?
This sample, along with SN 2014J, is limited to published cases SNe 1991T,
1998bu, 1995E, and 2006X.\footnote{References for SN max brightnesses and
$\tau_{sc}$: \citet{kri04}, \citet{her00}, \citet{rie99}, \citet{wae08}; for
echo brightnesses: \citet{qui06}, \citet{cap01}, \citet{cro08}, \citet{wan08}}
For quoted values of E(B-V) for these SNe, respectively: 0.1, 0.3, 0.7 and 1.4,
respectively, values of $V_{echo} - V_{SN-max}$ of 10.8, 10.3, 10.4 and 11.7
are found, with typical errors of about 0.1 mag on all of these quantities.
In comparison, Patat would predict $V_{echo} - V_{SN-max}$ values of about
10.8, 10.1, 10.5 and 11.7, respectively for these E(B-V) values, hence agreeing
typically to within 10-20\%.
In contrast, we measure a $V_{echo} - V_{SN-max}$ value for SN 2014J of 12.1,
corresponding to E(B-V) of 1.7 mag, or $\tau_{sc} = 4.1$ for $R_V$.
Since the these values are in the more heavily extincted domain, there errors
are relatively larger, about 20\% in $\tau_{sc}$.

The dark lane containing the projected position of SN 2014J is also largely
devoid of echo, at any radius from the SN.
Is this surprising?
For now, if we assume a simple slab model of dust for both the SN and dust in
the dark lane, we expect the echo surface brightness to follow a relation
well-approximated by $(1 - 10^{-0.4 A_B} ) \times 10^{-0.4 A_B}$, which rises
roughly linearly with extinction up to $A_B \approx 0.5$, and continues to a
peak surface brightness at $A_B \approx 1$, falling slowly to 10\% of the peak
surface brightness value at $A_B = 4$.
Thus the most the echo brightness in other regions could outshine the echoes
from the dust lane is 2.5 magnitudes.
Adding a synthetic echo signal in the dark lane at a strength 10\% of the
bright echo in the luminuos nebulosity east of the SN, this synthetic echo is
marginally detected.

The brightest nebulosity to the immediate East of the SN extends over radii of
5 pc to 15 pc from the SN, which the echo at $z \approx 330$ pc will traverse
by about April 2015, at which point it will enter another dark lane.
Similarly, echoes at this $z$ distance will enter other dark lanes over the
next few years.
There will be several opportunities to study these dark lanes before the close
of this decade.
(Note that for the most distant echoes seen, $z \approx 330$ pc, the distance
traveled perpendicular to the sightline to Earth is $\sim0.72
[t/(1 year)]^{1/2}$ arcsec.)

The environment around and in front of SN 2014J appears to be an interesting,
unprecedented case in terms of light echo environments.
These data appear to show an unfolding, rich interstellar environment, but one
that is likely to involve some significant but not prohiibitive complexity in
separating actual interstellar structure of dust clouds as traced by echoes, as
opposed to echo structure which is imposed by shadowing by extinction between
the echoing cloud and the SN, extinction which will in itself produce its own
echo signal.
While echoes from SNe 1987A, 1993J and 2006X may suffer modestly from such
effects (as seen in unpublished work by the authors), SN 2014J is probably a
clear case in which a large amount of interstellar structure can be mapped in
increasing refinement by collecting data over multiple epochs and then using
this to iteratively reconstruct not only the interstellar dust distribution
but how clouds are shadowing more distant clouds along the same sightline to
SN 2014J.
We are developing the techniques required to accomplish this, for SN 2014J and
other SNe.
Given this technique and sufficient data, one might even constrain the solid
angular distribution of extinction of circumstellar matter and the environment
so close to the SN as to have been missed by echo observations, in support of
other probes of near-SN space.

\section{Conclusions}

At least one and most likely two light echo signals are detected from SN 2014J
corresponding material about 330 pc and likely 80 pc in the SN foreground,
and the former is well-correlated in spatial extent with structures seen in
two-dimensional projection.
Initial three-dimensional maps of the former structure seems consistent with
two inclined planes at nearly the same distance in front of the SN, separated
by a dark lane.
The inner, probable echo appears likely to form only a small part of what might
be a more complex group of echoes at distance roughly 50 - 120 pc in the
foreground of SN 2014J, and future data on these other possible echo clouds may
alter the detailed interpretation of the structure of the more distant echoes
by providing more information about the effects of shadowing of light from
SN 2014J in this dense dust environment.
Future epochs of echo image are expected to clarify these and other issues
regarding circumstellar and interstellar structure around and in front of SN
2014J.

\acknowledgments

Based on observations made with the NASA/ESA Hubble Space Telescope,
obtained at the Space Telescope Science Institute, which is operated by the
Association of Universities for Research in Astronomy, Inc., under NASA
contract NAS 5-26555. These observations are associated with program \#13646.

\begin{figure}
\hskip -0.5in
\vskip -1.0in
\includegraphics[angle=90,scale=.70]{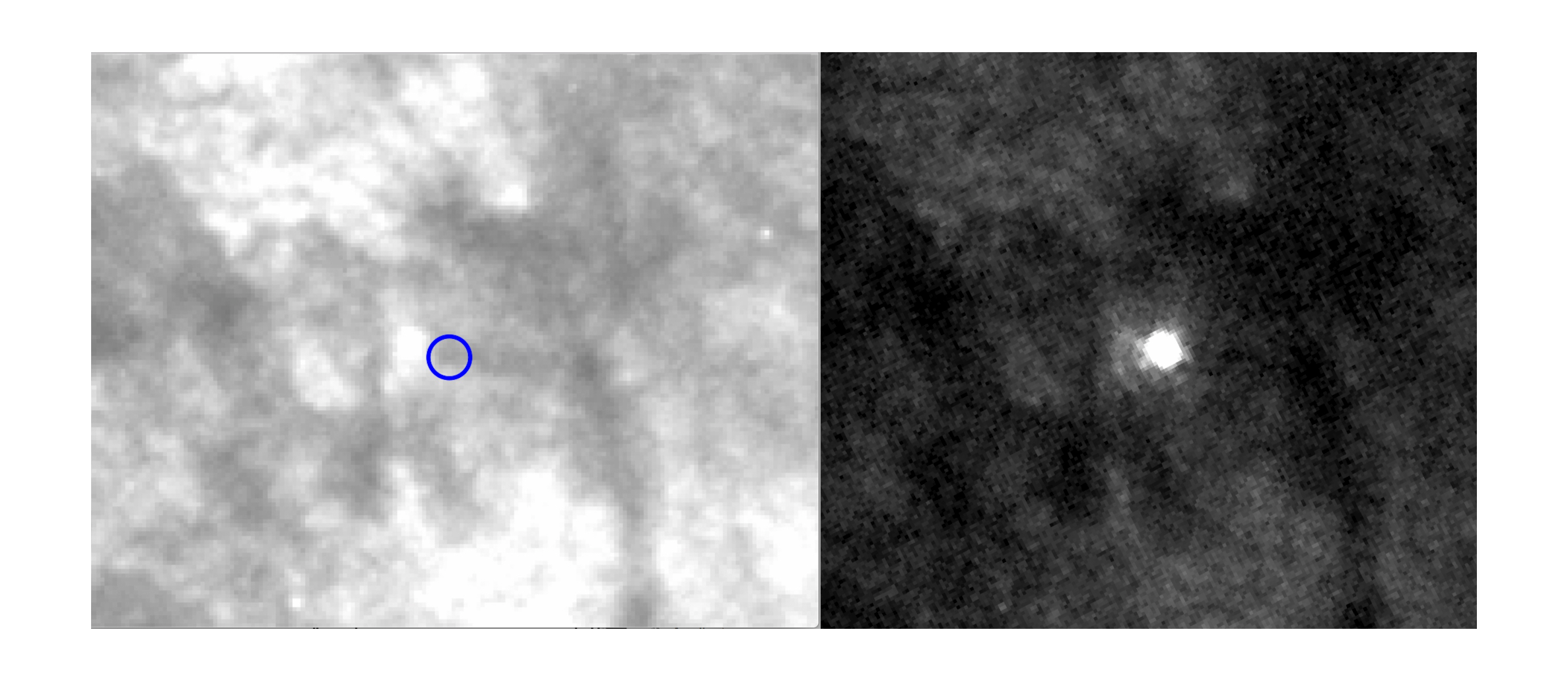}
\caption{
{\bf Left panel (a):}
View of M82, 0.9 kpc west of the galaxy's nucleus and centered on SN 2014J,
covering a field 8.4 arcsec = 143 pc across (North up, East left).
The image was taken in 2006 March 29 with $HST$/ACS(WFC) in the F435W band as
part of HST program \#10776.
The central circular mark corresponds to a 10 pc diameter, and is centered on
the SN.
Note that the SN is centered at one end of a dark lane, and sits just west of a
particularly bright patch of nebulosity.
{\bf Right panel (b):}
View of the same 143 pc field in M82 as in Fig. 1a, taken by $HST$/WFC3(UVIS) on
2014 September 5 (213 d after maximum) in F438W as part of program \#13626
(Crotts, PI).
SN 2014J is indicated as the bright point source in the center.
Note in addition the apparent nebulosity out to radii of about 12 pc,
especially just to the east of the SN.
(The nebulosity of M82 is less apparent than in Figure 1a due to the reduced
contrast in displaying these data.)\label{fig1}}
\end{figure}

\begin{figure}
\vskip -5.0in
\includegraphics[angle=00,scale=.70]{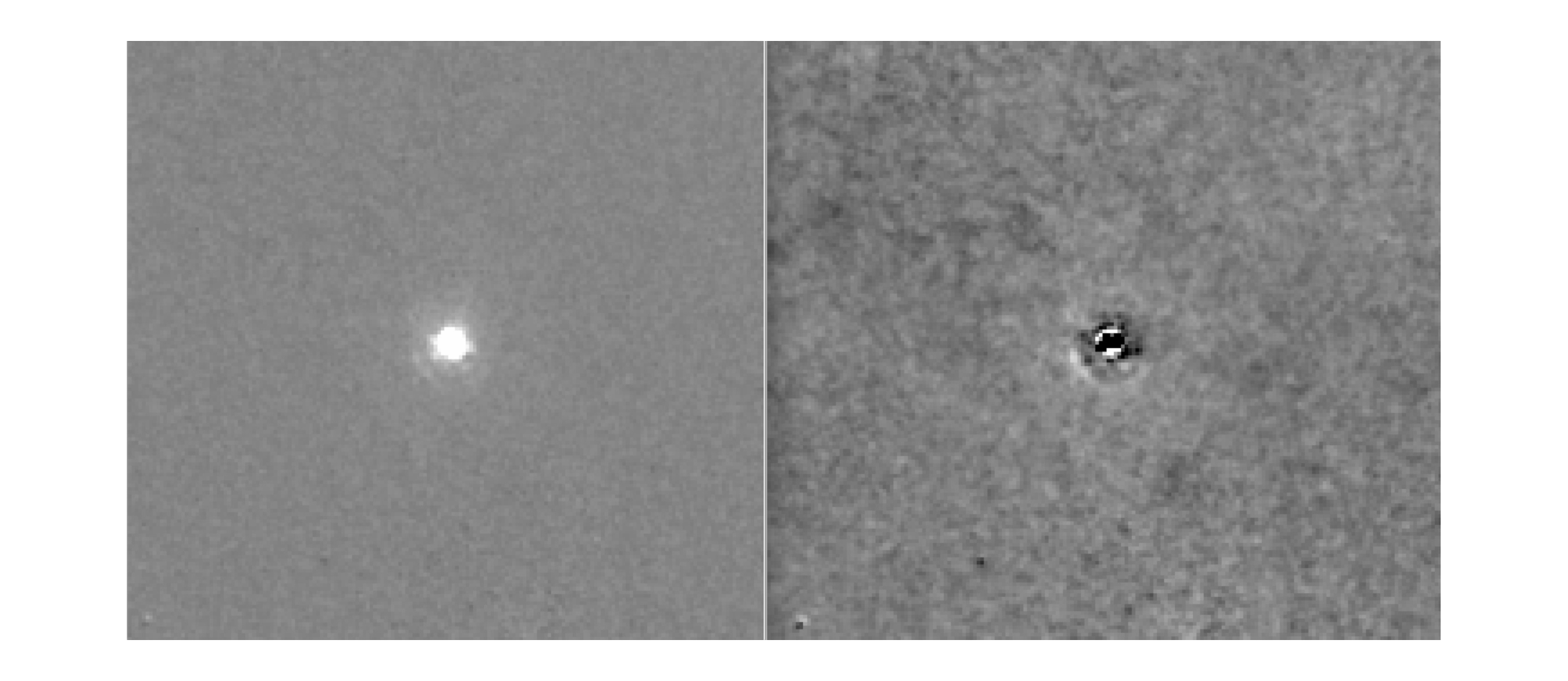}
\caption{
{\bf Left panel (a):}
The image in Figure 1a subtracted from that in Figure 1b, after being scaled in
flux and registered in spatial coordinates.
The background does not subtract perfectly, leaving a mottled appearance due to
color terms left from the F435W band in ACS(WFC) versus the F438W band in
WFC3(UVIS).
Note that even more apparent residual nebulosity has appeared with the 
explosion of SN 2014J, at radii of about 0.6 arcsec, faint but evident.
{\bf Right panel (b):}
The image in Figure 2a with the SN point source subtracted (as derived from the
8s F438W WFC3(UVIS) SN image from 35 d after maximum.
The echo ring at 10 - 13 pc is even more apparent, mainly south of the SN (at
PAs 85 to 230), and to a lesser extent due North (PAs -40 to 20), but not
significantly to the East and West (PAs 25 to 80, and 230-250 and 245-315,
respectively).
The echoes are largely absent from the area noted as the dark lane in Figure
1a, while it correlated well with bright nebulosity at other PAs.\label{fig2}}
\end{figure}

\begin{figure}
\hskip +0.0in
\includegraphics[angle=-90,scale=.60]{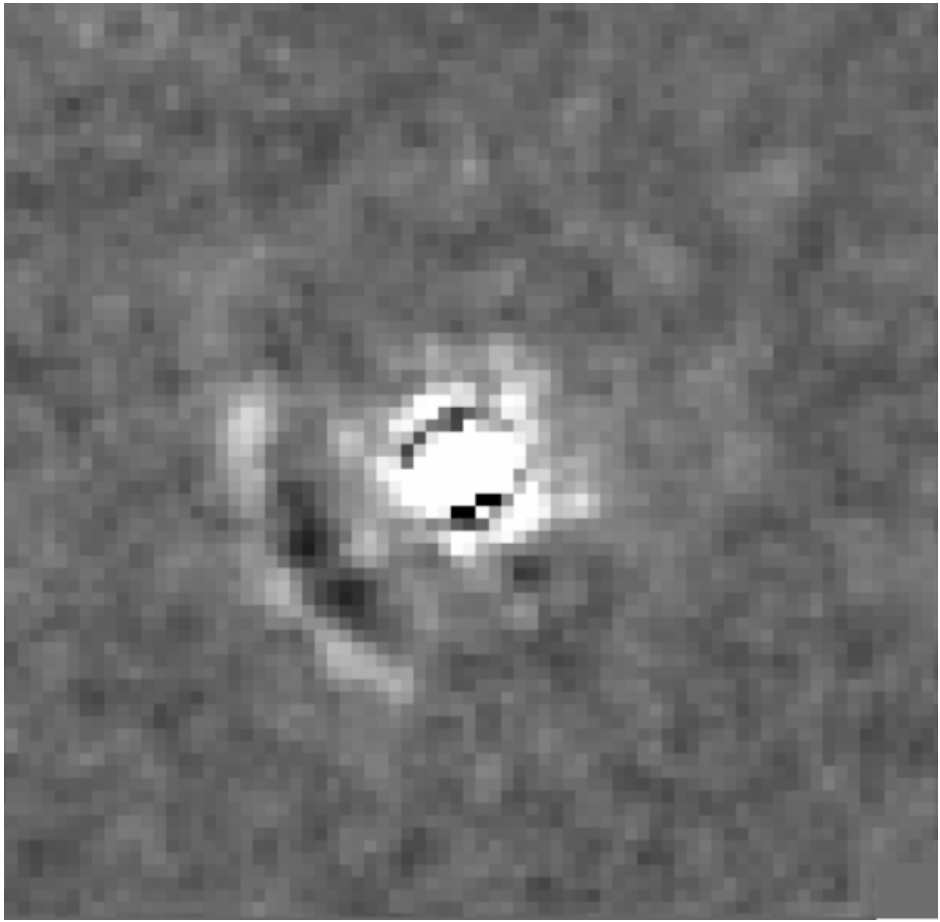}
\vskip +0.0in
\caption{
A 3-arcsec wide field centered on SN2014J shows the change in flux distribution
between 2014 September 5 and 2015 February 2 in the F438W band (with the later
epoch appearing bright, the earlier dark).
This readily reveals bright echoes as a positive signal in advance of a
negative one.
New echoes will appear as positive-only signals.
Where the 11 - 13 pc ring was strongest in 2014 September e.g., PA 85 to 170,
it is still strong.
At other PAs in this representation it is marginally detected.
Significantly, the hint of an inner echo at $r \approx 0.4$ arcsec, PA$\approx
215^\circ$ is confirmed; in fact this signal has spread to PA 205 to 225.
Additionally, strong positive-only signals have appeared at similar radii
$0.3$ arcsec $\la r \la 0.5$ arcsec, beyond the range of PSF
subtraction systematic errors, suggesting a complex of structures ranging over
about a factor of $\pm 40$\% in foreground distance $z$.}
\end{figure}

\begin{figure}
\hskip -0.5in
\includegraphics[angle=90,scale=.80]{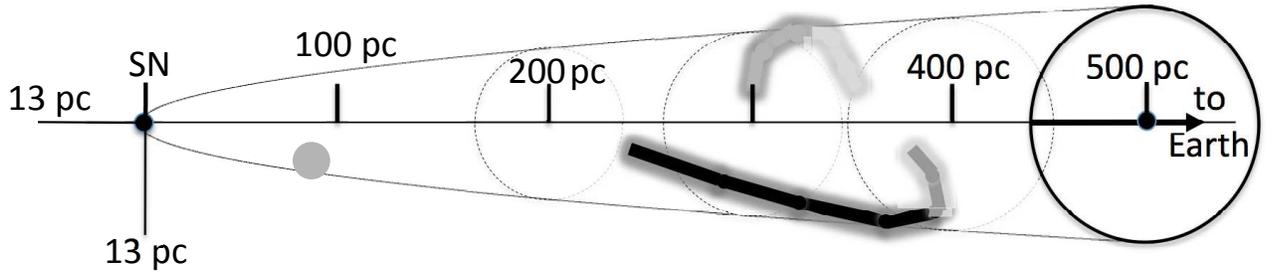}
\vskip -2.0in
\caption{
A cartoon view of the echo loci plotted on distorted orthonormal axes along
with the echo paraboloid for 213 d after SN 2014J's maximum light.
The view point is in a direction near that towards Earth, such that the
Earth-SN sightline (long horizontal line) is foreshortened by a factor of four.
The echo ring at about 12 pc radius is portrayed as the gray forms above and
below the sightline (N is up, E out of the page).
The density of the grayscale for these forms indicates the brightness of the
echo, with features behind the paraboloid also suppressed in grayscale.
Note the primary echo complex at a distance $z$ of about 330 pc in front of the
SN, which has the rough shape of two inclined surfaces, at large distance from
the SN due north and south, and smaller distance east and west, consistent with
the oval shape of the echo ring.
The candiate echo at smaller radii maps to about $z = 80$ pc.}
\end{figure}

\begin{figure}
\plottwo{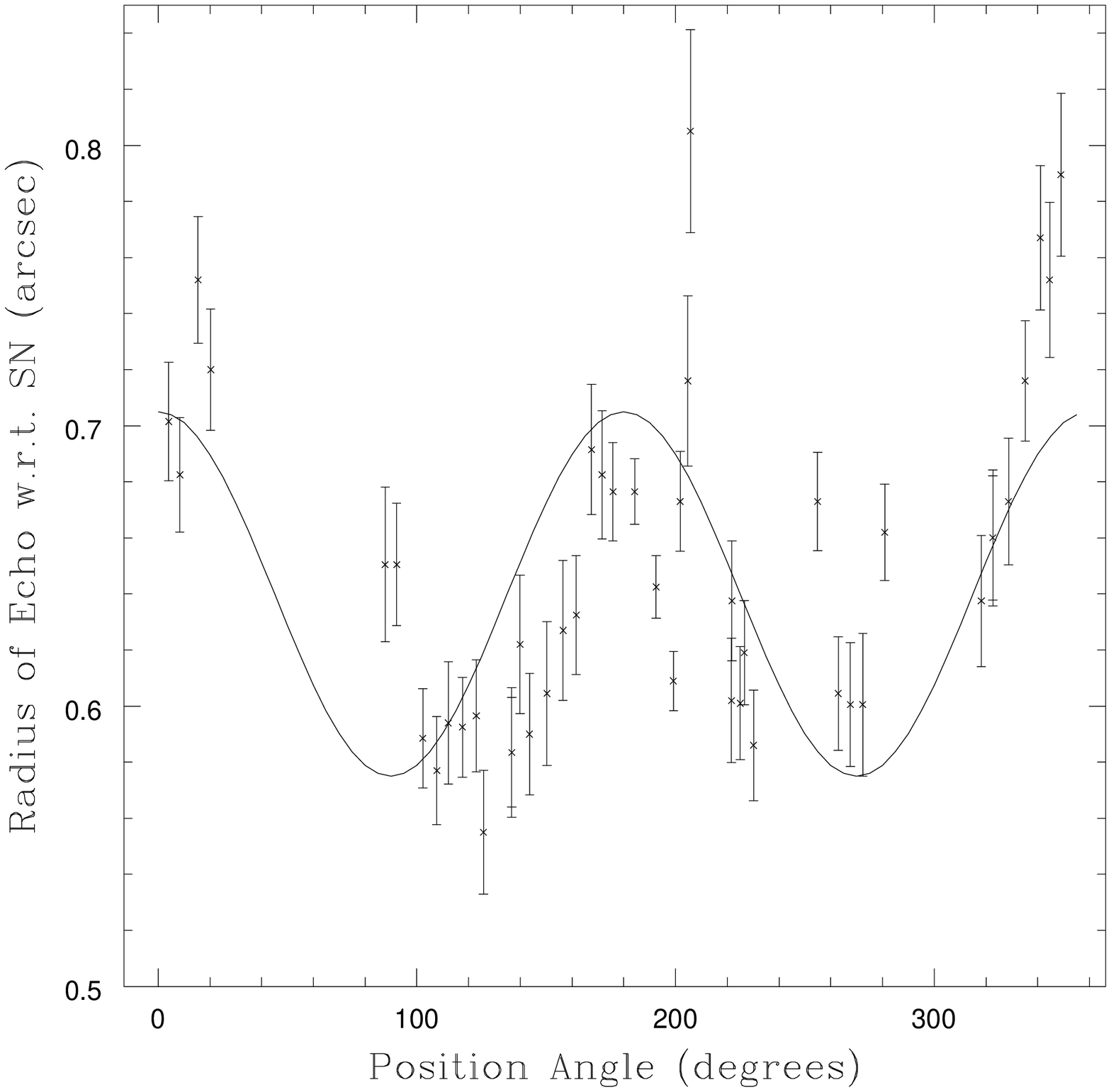}{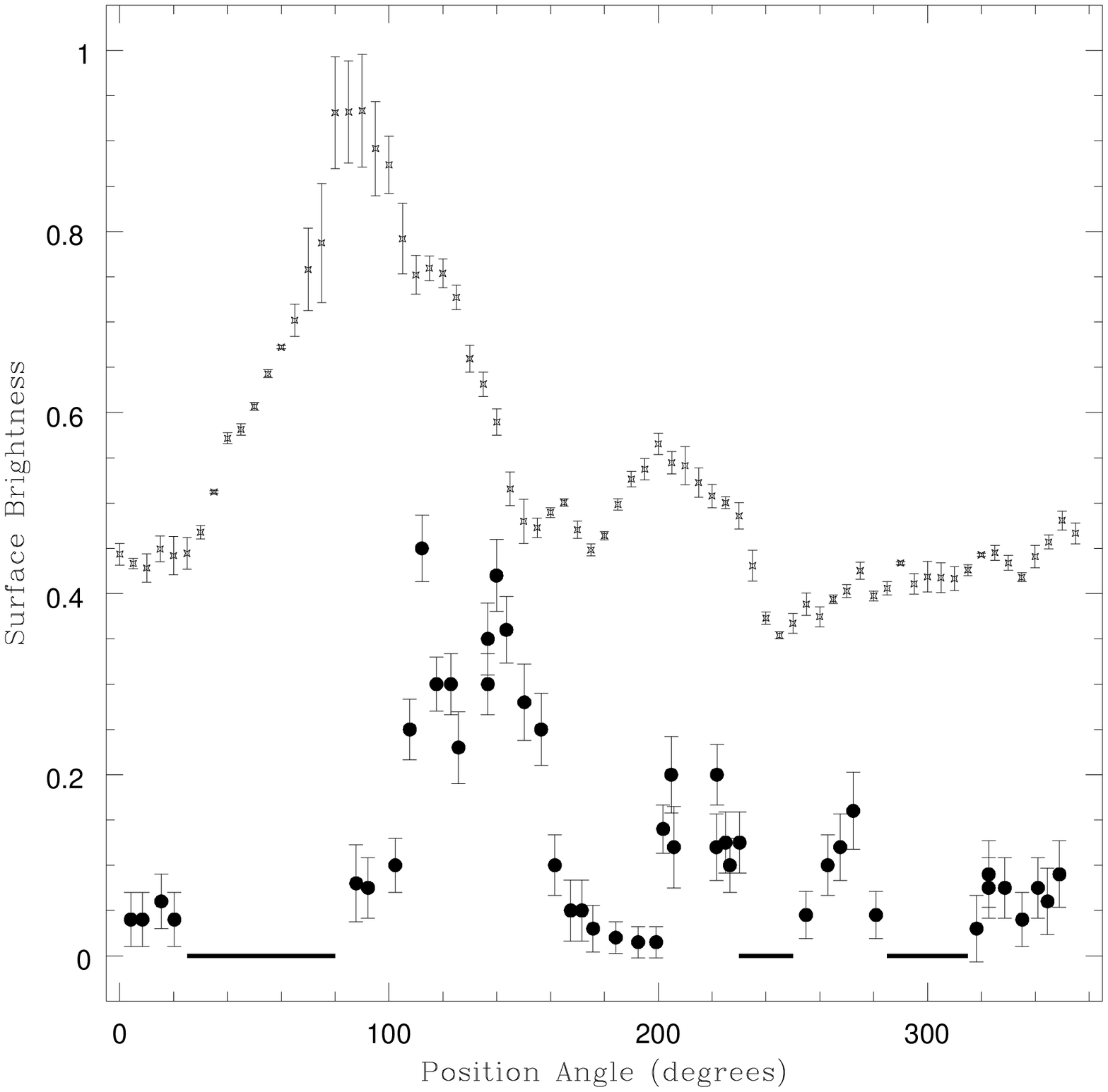}
\caption{
{\bf Left panel (a):}
Data points and errorbars represent the position of the primary echo around
SN 2014J as a function of position angle, as of 213 d after maximum light, plus
a best-fit ellipse centered on the SN point source.
{\bf Right panel (b):}
The surface brightness at the echo locus of the underlying M82 nebulosity
(small crosses) in the ACS/F435W band and light echo 213 d (large dots) after
maximum as seen in the WFC3/F438W band.
The echo surface brightness is not measureable (but consistent with zero) at
certain position angles where the width of the echo cannot be measured, as
indicated by bars at zero surface brightness near PA = 50, 240 and 300$^\circ$.
Surface brightnesses are in units of approximately $10^{-18}$ erg s$^{-1}$
cm$^{-2}$ \AA$^{-1}$ arcsec$^{-2}$.
}
\end{figure}

\end {document}